# The distribution of oxygen at the $Ni_{81}Fe_{19}$/Ta interface

T.J.A. Mori[a,b], R.D.D. Pace[a,c], W.H. Flores[b,d], M. Carara[a], L.F. Schelp[a], L.S. Dorneles[a,*]

[a] Universidade Federal de Santa Maria, Departamento de Física, Santa Maria, RS, Brazil, 97105-900.

[b] Laboratório Nacional de Luz Síncrotron, Centro Nacional de Pesquisa em Energia e Materiais, Campinas SP, Brazil, 13083-970.

[c] Universidade Federal do Rio Grande do Norte, Departamento de Física Teórica e Experimental, Grupo de Nanoestruturas Magnéticas e Semicondutoras, Natal, RN, Brazil, 59078-900.

[d] Universidade Federal do Pampa, Campus Bagé, Bagé, RS, Brazil, 96413-172.

[*] Corresponding author. Tel.: +55-55-3220-8618. Electronic mail: lucio.dorneles@ufsm.br (L.S. Dorneles).

## ORCID

T.J.A. Mori: https://orcid.org/0000-0001-5340-3282
R.D.D. Pace: https://orcid.org/0000-0003-4406-9512
W.H. Flores: https://orcid.org/0000-0002-3576-4751
M. Carara: https://orcid.org/0000-0002-9412-093X
L.S. Dorneles: https://orcid.org/0000-0001-5833-1775

## Abstract

The knowledge of how oxygen atoms are distributed at a magnetic-metal / oxide, or magnetic-metal / non-magnetic-metal interface, can be an useful tool to optimize device production. Multilayered $Ni_{81}Fe_{19}$/Ta samples consisting of 15 bilayers of 2.5 nm each, grown onto glass substrates by magnetron sputtering from $Ni_{81}Fe_{19}$ and Ta targets, have been investigated. X-ray absorption near edge structure, extended X-Ray absorption fine structure, small angle X-ray diffraction, and simulations, were used to characterize the samples. Oxygen atoms incorporated onto $Ni_{81}Fe_{19}$ films during $O_2$ exposition are mainly bonded to Fe atoms. This partial oxidation of the $Ni_{81}Fe_{19}$ surface works as a barrier to arriving Ta atoms, preventing intermixing at the $Ni_{81}Fe_{19}$/Ta interface. The reduction of the $Ni_{81}Fe_{19}$ surface by the formation of $TaO_x$ is observed.

# Introduction

In tunneling junctions the tunnel magnetoresistance (TMR) is mainly determined by the spin polarization at the magnetic surfaces touching the insulating barrier, and the presence of a nano-oxide-layer (NOL) at the metal/barrier interface decreases the TMR ratio. In multilayered spin valves the giant magnetoresistance (GMR) ratio can be improved by the insertion of a NOL at the interfaces [1,2]. This improvement is also observed in magnetic field sensors, where the anisotropic magnetoresistance (AMR) ratio can be enhanced by the insertion of a NOL at the interfaces, by modifying the interface flatness and the specular reflection of the conduction electrons [3].

At thermodynamic equilibrium, the distribution of oxygen atoms at a metal/metal interface will be determined by the difference in oxidation energy of the metals. During the deposition of very electronegative materials like Ta onto another metal's surface, the reduction of this surface can occur if it presented some initial degree of oxidation [4]. This capability of extracting oxygen from an oxidized surface would explain why the interface can work as an anti-diffusion barrier for oxygen atoms. But, during thin film deposition, oxides may not have a well defined composition, so energy considerations work as a guide.

The direct measurement of oxygen states is not simple and has been restricted to a few studies with X-ray photoelectron spectroscopy [5], real time resistivity measurements, or resonant scattering [6]. Results in X-ray absorption fine structure (XAFS) measurements from thermally annealed CoFe/AlO$_x$ junctions [7] have shown that the oxygen transfer from the magnetic material to the insulating barrier can be activated by annealing.

In tunneling junctions, the insulating barrier can be produced by the oxidation of a metal film, exposed to a natural or plasma assisted oxidation process. The interface prevents the oxidation of the bottom electrode and the insulator has a higher oxygen content than in the more stable stoichiometry [8]. Without this reduction mechanism it would be hardly justifiable even the existence of TMR in tunneling junctions produced using *ex-situ* masks exchange; the presence of oxygen atoms at the surface of a magnetic film, even if restricted to one or two atomic monolayers, would introduce spin-independent current channels eroding the TMR ratio [2].

In magnetic sensors, the presence of a NOL allows for the annealing-induced improvement of the multilayered microstructure; the NOL acts as an interdiffusion barrier and as an oxygen-reduction agent [3].

The knowledge of how oxygen atoms are distributed at a metal/metal interface can be an useful tool to optimize device production. In this work we have investigated the $Ni_{81}Fe_{19}$/Ta interface in multilayered samples produced under a controlled atmosphere, where we have deliberately introduced $O_2$ at different stages of the sample preparation process.

# Experimental Details

In order to measure XAFS signal-to-noise ratios sensitive to the oxidation at the interfaces, multilayered samples of very thin $Ni_{81}Fe_{19}$/Ta layers were produced. The multilayers consisted of 15 bilayers of 2.5 nm each. The films were grown onto glass substrates from $Ni_{81}Fe_{19}$ (Py) and Ta targets using RF or DC magnetron sputtering, respectively, in a 0.7 Pa argon atmosphere. The thickness of the Py and Ta layers were fixed at 2.0 and 0.5 nm, respectively. Ta was chosen because it presents an absorption edge that could be detected under our experimental conditions.

The sample PyTa was grown without any deliberate exposition to oxygen during growth. The samples PyO2Ta and PyTaO2 were grown with an additional step; oxygen was admitted into the deposition chamber (13.3 Pa for 300 s after each Py or Ta layer, respectively), and then the chamber was pumped back down to base pressure (~ $2x10^{-8}$ Pa) before the growth of the next layer. The sample PyO2 was grown with the same exposition to oxygen after the deposition of each Py layer, and with no deposition of Ta. All samples were exposed to ambient atmosphere after growth.

The chemical modulation perpendicular to the substrate surface was observed using small angle X-ray reflectometry (XRR). In order to obtain a proper standards for comparison, Py and $Ta_2O_5$ films were grown at room temperature both onto Si or glass substrates. The interference patterns were acquired at small angles $\theta$ within the 0.25° to 2.50° range, using a Brüker AXS D8 Advance diffractometer in Bragg–Brentano geometry ($\theta$-$2\theta$) equipped with Göbel mirrors to generate a parallel beam. Simulations of the experimental data were performed in order to obtain further informations such as thicknesses, roughnesses and densities of each sample, using the "Software for modeling the optical properties of multilayered films - IMD", whose methods and calculations are outlined in detail in reference [9].

The XAFS experiments were performed at the XAFS1 beam-line of the Laboratório Nacional de Luz Síncrotron, near room temperature. XANES and EXAFS of the samples were measured at the $K$ absorption edge of Ni and Fe using a 15-elements Ge fluorescence detector, and at the $L_3$ absorption edge of Ta using partial electron yield mode. Standard metal or oxide films were also measured under the same conditions, and used for comparison. At least four scans were collected from each multilayer. The average of the normalized spectra were analyzed following the standard procedure described in reference [10]. Background was removed using a fitted cubic spline curve. The EXAFS oscillations were then Fourier-transformed (FT) using weight 2, and a Hanning window ranging from 2 to 12 Å$^{-1}$ in $k$-space. The distances shown in the FT were not phase-shift-corrected. The data was fitted in *k*-space after Fourier back-transform of the first metal-metal peak.

# Results

Fig. 1 shows the Fe (7112 eV) and Ni (8333 eV) *K*-edge XANES spectra, normalized, for all the samples and the standards. XANES spectra shapes depend essentially on the coordination geometry around the absorbing atom [11]. The position and the intensity of these features are typical of metal-metal or of a metal-oxygen coordination. The metal-oxygen coordination exhibits a very high resonance at the edge (called white line) due to a dipole transition ($1s \rightarrow 4p$), which indicates a high concentration of unoccupied states that is typical of oxides.

It can be clearly seen from the spectrum from sample PyO2 that the Py exposition to the oxygen atmosphere led to oxidation. The oxidation signatures are present Fe and Ni absorption edges, and are much more evident in the Fe spectrum in both Fig.1(a) (at the absorption edge, indicated by an arrow) and Fig.1(b) (just after the absorption edge). For sample PyO2 it is very clear a significant Fe-O bond contribution, and the white line is not so pronounced in the spectra from the other multilayers, nor in the Ni *K*-edge spectra. Considering the Fe/Ni atomic ratio of ~ 1/4, this is a clear evidence that the oxygen atoms tend to bond to the iron atoms as expected, according to the oxidation energy associated with Fe-O (-68.1 kJ/mol) or Ni-O (-50,6 kJ/mol) bonds.

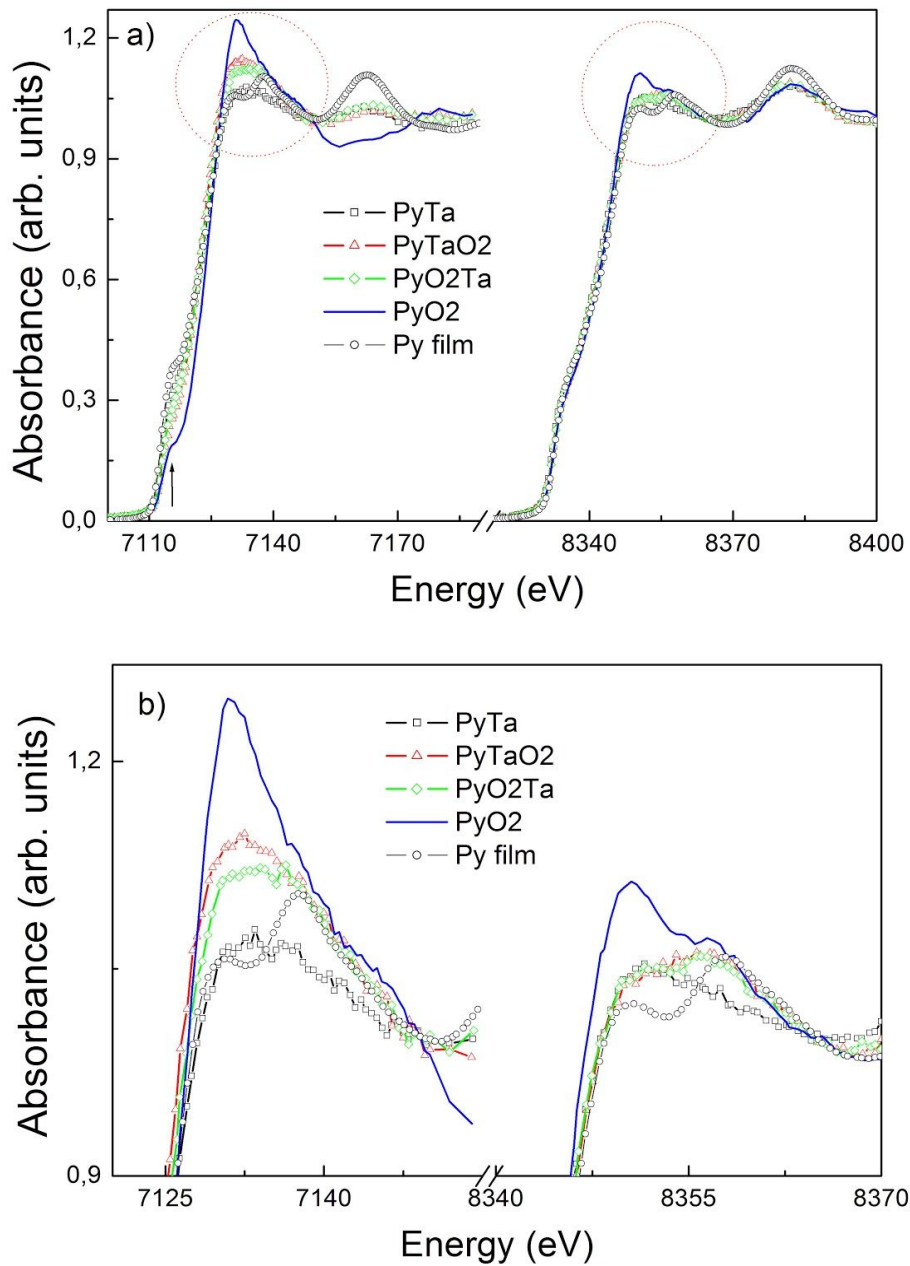

Fig. 1. (a) Normalized Fe (left) and Ni (right) *K*-edge XANES spectra from the multilayers. A spectrum from a Py film is presented for comparison. (b) XANES region of the main resonance (white line region) on the expanded energy scale indicated in (a).

The growth of the Ta layer on top of the Py layer produces some degree of disorder, as can be seen in the spectrum from sample PyTa; the degree of disorder can be inferred from the Debye-Waller factors shown in the last rows of Table 1, higher for this sample in both edges. It can also be inferred from the shape of the peak observed in the post-edge region [12], as shown in Fig.1(b). This degree of disorder can be related to the kinetic energy of the impinging Ta atoms leading to intermixing or alloying at the interface. As will be discussed later, the differences between samples PyO2Ta and PyTaO2 seem to be linked to this degree of disorder.

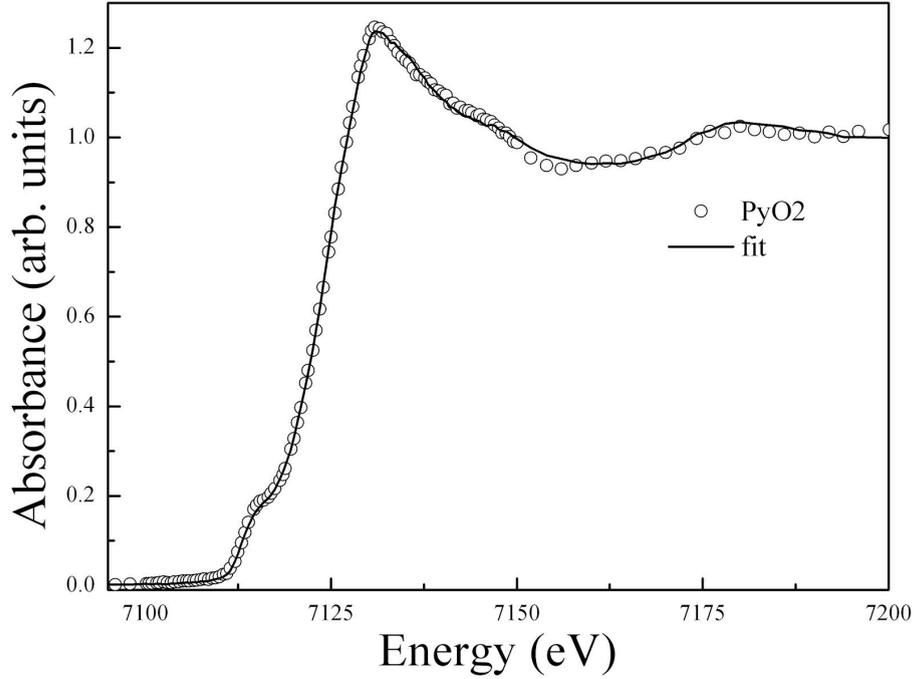

Fig. 2. Fe *K*-edge XANES spectrum from PyO2 (open circles) and a linear combination of spectra from a $Fe_2O_3$ and a Py film (line).

Table 1: XAFS analysis results using WinXas program [11]. XANES: weight percent contribution from oxide ($Fe_2O_3$ or NiO) and metal (Py) phases to the measured spectra of the Fe and Ni edge. EXAFS: the Debye-Waller factors ($\sigma^2$) are given $\times 10^{-3}$ Å$^2$.

| Sample | $Fe_2O_3$ (wt. %) | Py (wt. %) | NiO (wt. %) | Py (wt. %) | Disorder Fe-edge $\sigma^2$ ($10^{-3}$ Å$^2$) | Disorder Ni-edge $\sigma^2$ ($10^{-3}$ Å$^2$) |
|---|---|---|---|---|---|---|
| **PyTa** | 11.22 ± 0.43 | 88.78 ± 0.47 | 2.82 ± 0.04 | 97.18 ± 0.04 | 8.32 ± 0.098 | 5.28 ± 0.078 |
| **PyTaO2** | 28.51 ± 0.05 | 71.49 ± 0.06 | 1.18 ± 0.04 | 98.82 ± 0.42 | 5.40 ± 0.017 | 5.16 ± 0.015 |
| **PyO2Ta** | 20.29 ± 0.52 | 79.71 ± 0.57 | 1.68 ± 0.04 | 98.32 ± 0.04 | 4.60 ± 0.019 | 3.66 ± 0.070 |
| **PyO2** | 50.24 ± 0.30 | 49.76 ± 0.32 | 6.13 ± 0.04 | 93.87 ± 0.04 | 4.28 ± 0.020 | 3.31 ± 0.017 |

In samples PyO2Ta and PyTaO2 the Ta layers clearly extract oxygen atoms from the Py layers. To estimate the degree of oxidation in the multilayers, we can fit a calculated spectrum (obtained by the linear combination of spectra from oxide and metal phases) to the experimental spectrum, obtaining the $Fe_2O_3$, NiO, and Py weight contribution [13]. Table 1 shows the values obtained from this fitting procedure for all samples, and the Fe *K*-edge spectra from sample PyO2 are displayed in Fig. 2 as an example. From the values in Table 1 we can observe the trend of reduction in the oxidation level of the Py layer when Ta is introduced.

Fig. 3 shows (a) the EXAFS oscillations and (b) the FT at the Fe *K*-edge from all samples and from a Py film. The EXAFS oscillations are strongly damped with increasing *k*, characteristic of scattering by light atoms, oxygen in the present case. In Fig.3(a), the reduced amplitude (when compared to the Py film's amplitude) of the oscillations suggests disorder around the Fe atoms; and in the spectrum from sample PyO2 a contribution from Fe-O distance (indicated by the arrow) is also present.

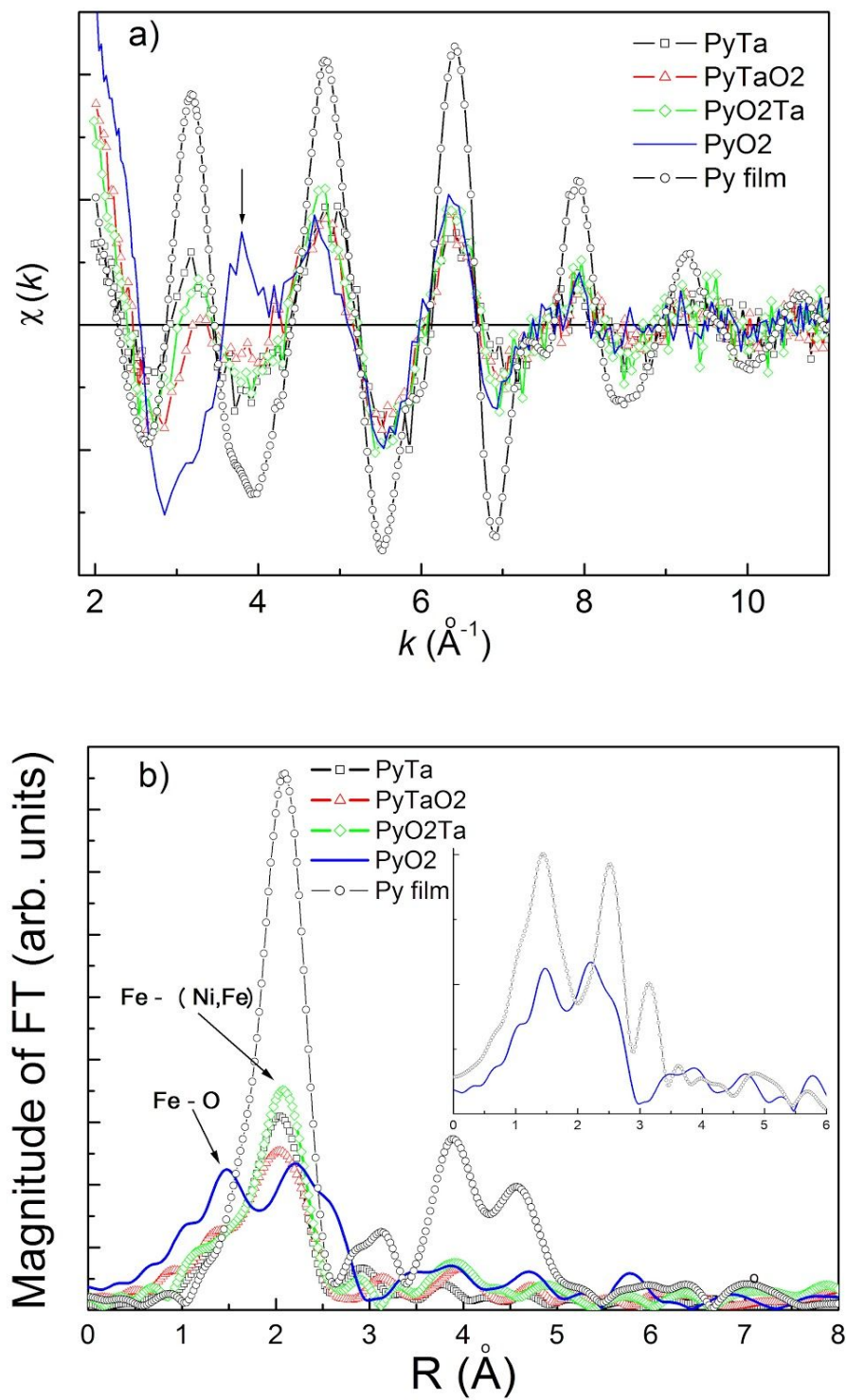

Fig. 3. (a) Fe *K*-edge EXAFS oscillations $\chi(k)$ from the multilayers and from the Py film. (b) FT (not corrected for phase shift) magnitude of $k^2$-weighted Fe *K*-edge EXAFS oscillations. The contributions from Fe-O and Fe-(Fe,Ni) are indicated in the figure. The inset shows the FT from the PyO2 film and a $Fe_2O_3$ powder sample.

The FT spectra shown in Fig. 3(b) are dominated by a peak centered around 2 Å (not corrected for phase shift) due to a Fe-(Fe,Ni) bond. The reduced magnitude of this peak, and the absence of more Fe-(Fe,Ni) peaks, are indicative of disorder. For the sample PyO2, the first observed peak is characteristic of a Fe-O bond. This can be better seen in the inset where the spectrum from $Fe_2O_3$ powder is plotted for comparison. For the other three samples, the first clear peak matches the peak for a Fe-(Fe,Ni) bond. And from the magnitude of these peaks for first neighbors, we can infer that Ta is effective in taking oxygen from the Py layers.

In accordance with what was indicated in the XANES measurement, the degree of disorder in the Py layer is lower in sample PyO2Ta than in samples PyTa or PyTaO2, where Ta is grown directly onto the metallic Py surface. It has been assumed that all metal atoms show a fix coordination number for the first shell (first peak in the FT), thus disorder is the key feature to distinguish between different deposition processes, shown in the last columns of Table 1. We have here the same behavior observed in the work from reference [3], where Al is deposited onto an oxidized Py surface that works as a barrier to the intermixing of the metallic layers.

The deposition of Ta onto the metallic Py surface leads to a defective interface, as indicated by the last two columns in Table 1, where disorder is higher for samples PyTa and PyTaO2; and this intermixing has an effect on the degree of oxidation of the samples. As shown in the first two columns of Table 1, obtained from measurements at the Fe $K$-edge, the $Fe_2O_3$ contribution is higher for sample PyTaO2 than for sample PyO2Ta. Also, as expected, the $Fe_2O_3$ contribution in these two samples is within the limits determined by the samples without $O_2$ exposition (PyTa) and without Ta deposition (PyO2).

Fig. 4 shows (a) the EXAFS oscillations and (b) the FT (not corrected for phase shift) of the Ni $K$-edge from all samples and the Py film. The oscillations at higher $k$ seen in Fig.4(a) clearly indicate the contribution from a Ni-(Fe,Ni) bond. The higher amplitude indicates a reduced disorder around the Ni atoms, corroborated by the observation of higher shells in the FT shown in Fig.4(b). The exposition to oxygen does not affect Ni as it does affect Fe, shown by the magnitude of the peak for first neighbors: all samples exposed to oxygen show a similar behavior. The magnitude of the peak for sample PyTa is a consequence of the absence of oxygen exposition, and as a consequence a higher degree of disorder at the interfaces.

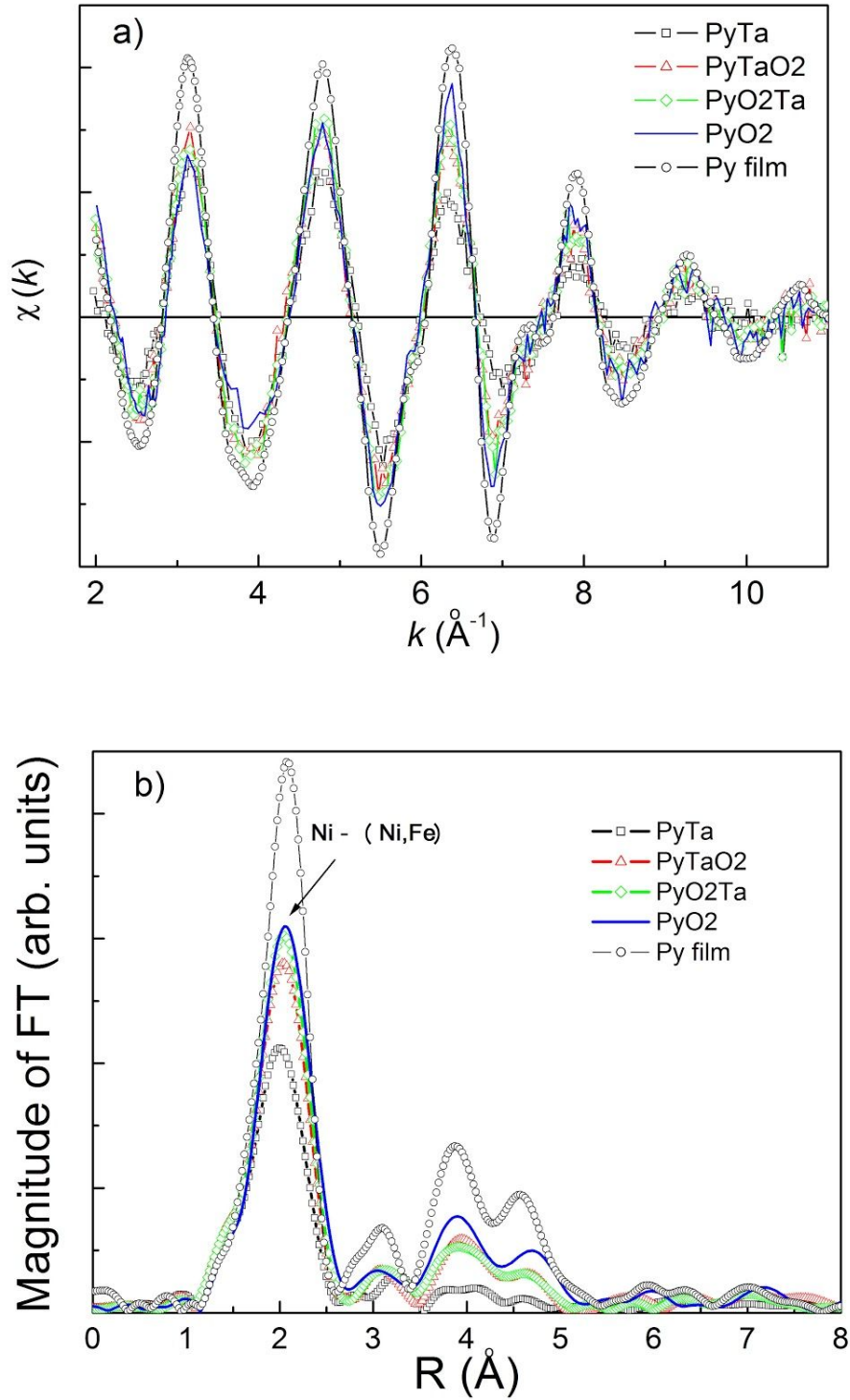

Fig. 4. (a) Ni *K*-edge EXAFS oscillations χ(k) from the multilayers and the Py film. (b) FT (not corrected for phase shift) magnitude of $k^2$-weighted Ni *K*-edge EXAFS oscillations. The contribution from Ni-(Ni,Fe) is indicated in the figure.

Fig.5 shows the Ta (9881 eV) $L_3$ XANES spectra for the multilayers, a Ta foil, and a $Ta_2O_5$ film. All spectra from the multilayered samples depart from that of the Ta metal, including sample PyTa as expected considering the results displayed on Table 1. As the thickness of the Ta layer corresponds to approximately 2 atomic monolayers, even atomically flat layers would not be completely surrounded by Ta atoms, and here we have also the presence of defects and/or intermixing. The intensity of the peaks in the spectra from samples PyTaO2 and PyO2Ta are close to the intensity of the peak in the spectrum from the $Ta_2O_5$ film, in accordance with what was previously observed in the Fe and Ni absorption spectra. We observe that reduction is more effective when Ta is deposited onto a previously oxidized Py surface.

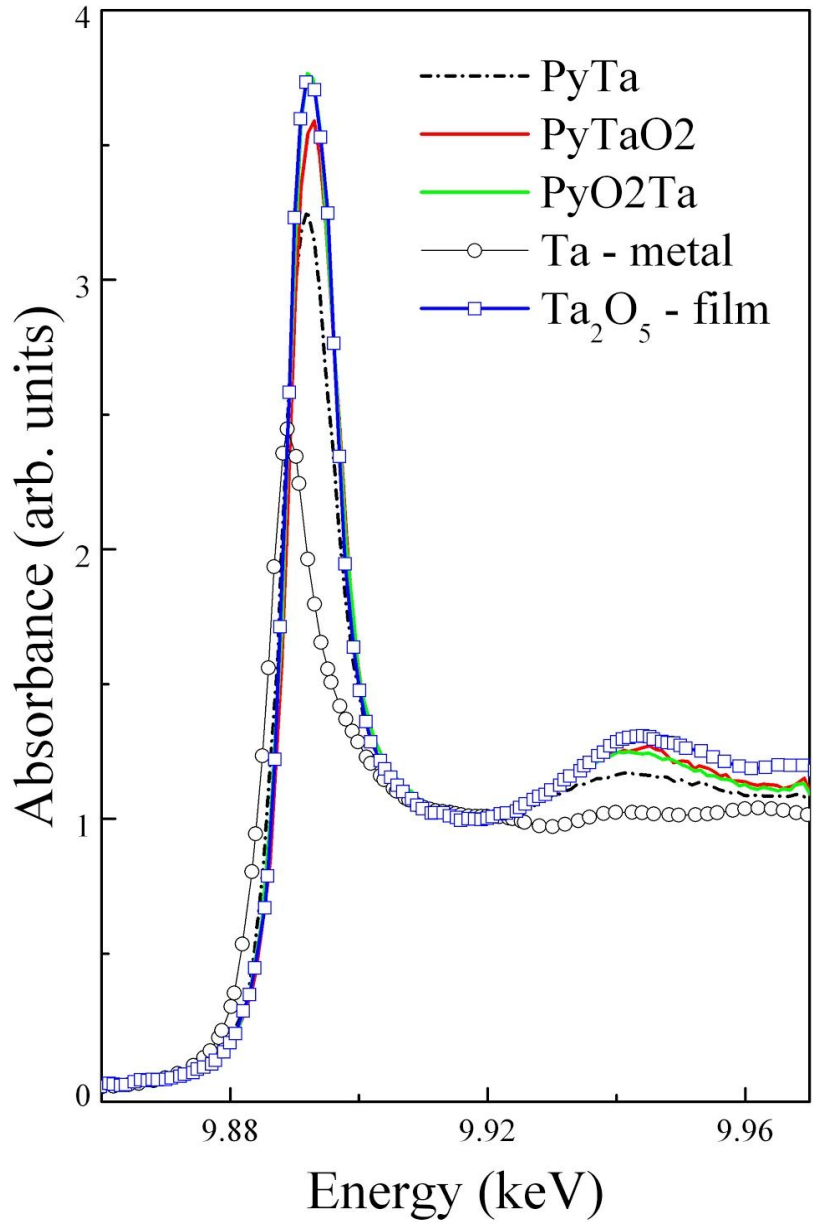

Fig. 5. Normalized Ta $L_3$-edge XANES spectra from the multilayers, a Ta metal foil, and a $Ta_2O_5$ film.

Fig.6 to Fig.9 show the XRR measurements and simulations using the values from Table 2. The simulations were made using a simple model: each multilayers is composed of a superposition of bilayers, and each layer has a different density and a fixed roughness. It is clear the chemical modulation perpendicular to the substrate surface, and the number of layers in each sample. The simulation is satisfactory up to relatively high angles $2\theta$ around 3°, but this simple model can not simulate the details of the measurement at higher angles; this is expressed by the error bars in Table 2.

The roughness of the PyO2Ta sample is the smallest, in accordance with the absorption experiments. The degree of disorder in the Py layer is lower when it is exposed to oxygen prior to the deposition of Ta, than when Ta is grown directly onto the metallic Py surface.

Table 2: XRR analysis results using IMD program [9].
*Means $PyO_x$ in the last line of the table.

| **Sample** | Density Py (g/cm³) | Density Ta (g/cm³) | Roughness Py (nm) | Roughness Ta (nm) | Thickness Py (nm) | Thickness Ta (nm) |
|---|---|---|---|---|---|---|
| **PyTa** | 9 ± 1 | 17 ± 1 | 1 ± 1 | 0.7 ± 0.1 | 2.0 ± 0.1 | 0.5 ± 0.1 |
| **PyTaO2** | 9 ± 1 | 7 ± 1 | 0.7 ± 0.1 | 0.4 ± 0.1 | 2.0 ± 0.1 | 1.0 ± 0.1 |
| **PyO2Ta** | 9 ± 1 | 8 ± 1 | 0.3 ± 0.1 | 0.3 ± 0.1 | 2.0 ± 0.1 | 0.9 ± 0.1 |
| **PyO2** | 9 ± 1 | 4 ± 1* | 0.8 ± 0.1 | 0.8 ± 0.1* | 1.7 ± 0.1 | 0.4 ± 0.1* |

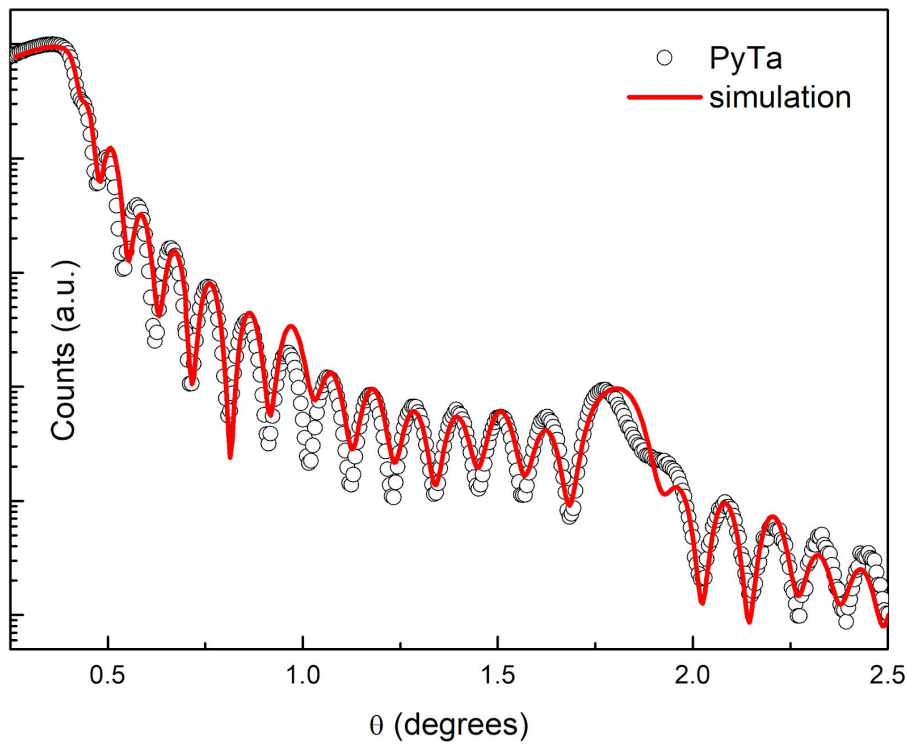

Fig.6. XRR measurement from sample PyTa and simulation.

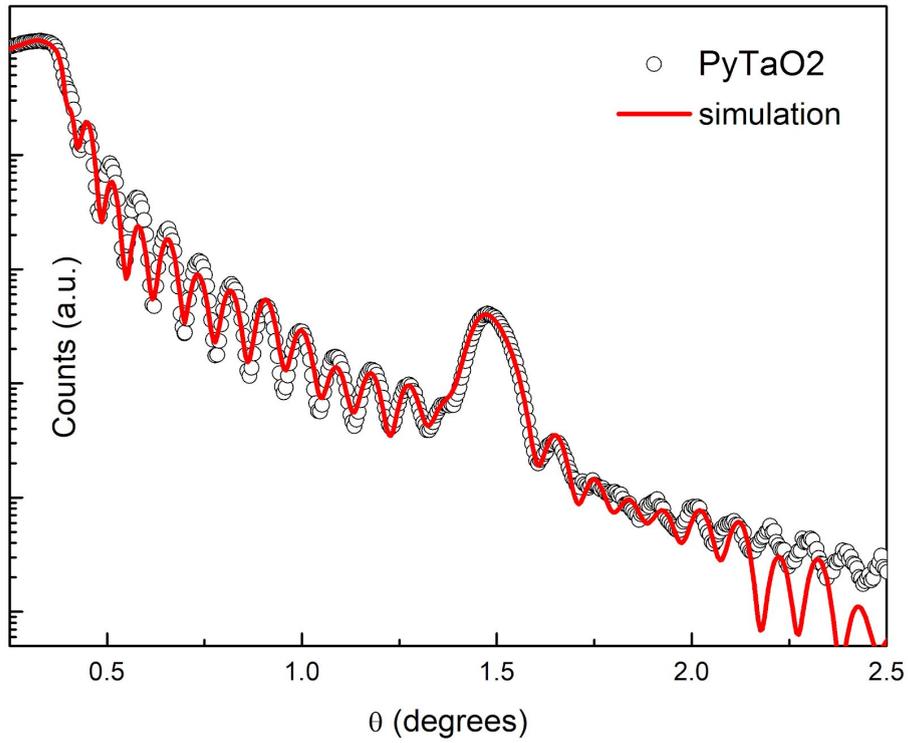

Fig.7. XRR measurement from sample PyTaO2 and simulation.

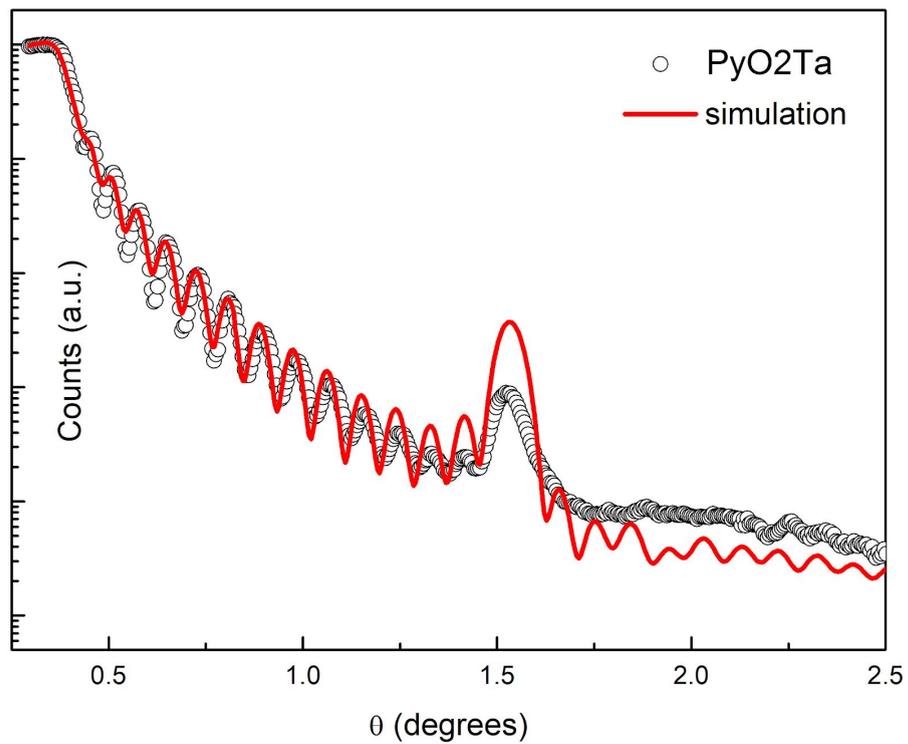

Fig.8. XRR measurement from sample PyO2Ta and simulation.

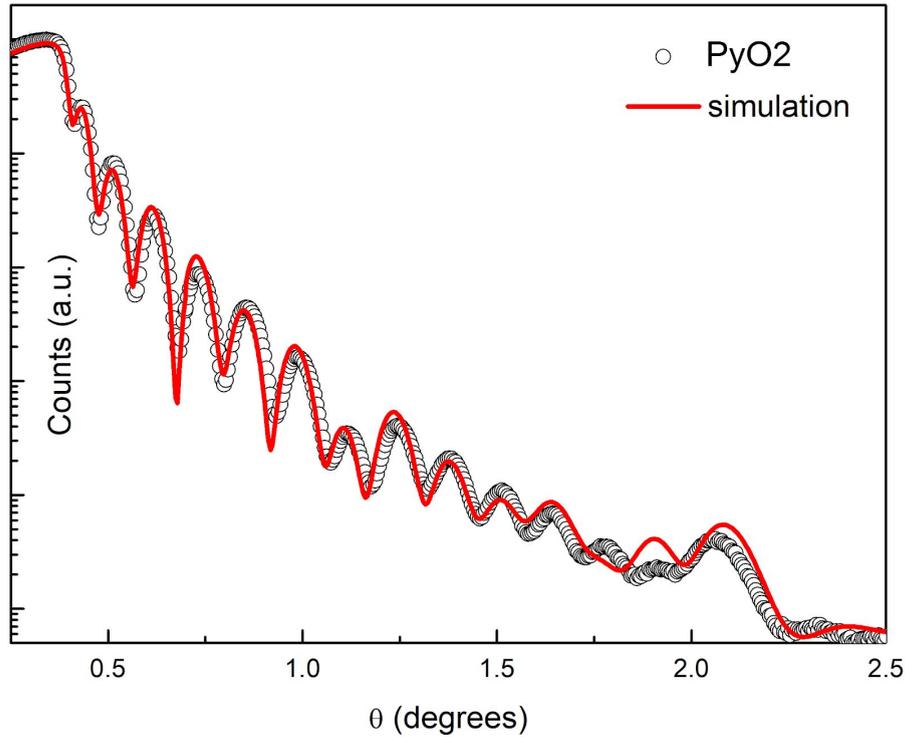

Fig.9. XRR measurement from sample PyO2 and simulation.

## Conclusions

We studied the oxygen distribution at $Ni_{81}Fe_{19}$ / Ta interfaces. The exposition of a $Ni_{81}Fe_{19}$ surface to an $O_2$ atmosphere bonds oxygen atoms mainly to Fe atoms. This oxide layer limits intermixing during the deposition of a very thin Ta film, acting as a barrier to the diffusion of Ta atoms into the $Ni_{81}Fe_{19}$ layer; during Ta deposition oxygen atoms are extracted from the $Ni_{81}Fe_{19}$ surface to form $TaO_x$. On the other hand, when the Ta film is deposited onto a $Ni_{81}Fe_{19}$ surface that has not been deliberately exposed to oxygen, intermixing or alloying takes place; this $Ni_{81}Fe_{19}$ / Ta interface layer is less effective in incorporating oxygen from the atmosphere.

## Acknowledgments

This study was financed in part by the Coordenação de Aperfeiçoamento de Pessoal de Nível Superior - Brasil (CAPES) - Finance Code 001, FAPESP, Laboratório Nacional de Luz Síncrotron (Brazilian Synchrotron Light Laboratory - LNLS), Brazil, and FAPERGS (grant PRONEX 2014). L.S.D. acknowledges financial support from CNPq (grant no. 302950/2017-6).